\documentclass[aip,jcp,amsmath,amssymb,preprint]{revtex4-1}

\usepackage{dcolumn} 
\usepackage{bm}
\usepackage{graphicx}
\usepackage{longtable}
\usepackage{footnote}
\usepackage{float}

\usepackage{color}

\draft 

\begin{document}

\newcommand{\bra}[1]{\langle #1|}
\newcommand{\ket}[1]{|#1\rangle}
\newcommand{\braket}[2]{\langle #1|#2\rangle}
\newcommand{\p}{^\prime}
\newcommand{\pp}{^{\prime\prime}}

\title{A highly accurate \textit{ab initio} potential energy surface for methane}

\author{Alec Owens} 
 \email{alec.owens.13@ucl.ac.uk}
\affiliation{Department of Physics and Astronomy, University College London, Gower Street, WC1E 6BT London, United Kingdom}
\affiliation{Max-Planck-Institut f\"{u}r Kohlenforschung, Kaiser-Wilhelm-Platz 1, 45470 M\"{u}lheim an der Ruhr, Germany}

\author{Sergei N. Yurchenko}
\affiliation{Department of Physics and Astronomy, University College London, Gower Street, WC1E 6BT London, United Kingdom}

\author{Andrey Yachmenev}
\affiliation{Department of Physics and Astronomy, University College London, Gower Street, WC1E 6BT London, United Kingdom}

\author{Jonathan Tennyson}
\affiliation{Department of Physics and Astronomy, University College London, Gower Street, WC1E 6BT London, United Kingdom}

\author{Walter Thiel}
\affiliation{Max-Planck-Institut f\"{u}r Kohlenforschung, Kaiser-Wilhelm-Platz 1, 45470 M\"{u}lheim an der Ruhr, Germany}

\date{\today}

\begin{abstract}
A new nine-dimensional potential energy surface (PES) for methane has been generated using state-of-the-art \textit{ab initio} theory. The PES is based on explicitly correlated coupled cluster calculations with extrapolation to the complete basis set limit and incorporates a range of higher-level additive energy corrections. These include: core-valence electron correlation, higher-order coupled cluster terms beyond perturbative triples, scalar relativistic effects and the diagonal Born-Oppenheimer correction. Sub-wavenumber accuracy is achieved for the majority of experimentally known vibrational energy levels with the four fundamentals of $^{12}$CH$_4$ reproduced with a root-mean-square error of $0.70{\,}$cm$^{-1}$. The computed \textit{ab initio} equilibrium C{--}H bond length is in excellent agreement with previous values despite pure rotational energies displaying minor systematic errors as $J$ (rotational excitation) increases. It is shown that these errors can be significantly reduced by adjusting the equilibrium geometry. The PES represents the most accurate \textit{ab initio} surface to date and will serve as a good starting point for empirical refinement.
\end{abstract}


\maketitle 

\section{Introduction}
\label{sec:intro} 

 As a key atmospheric molecule the infrared spectrum of methane (CH$_4$) has been the subject of numerous studies. Its complex polyad structure is beginning to be explored in greater detail at higher energies,~\citep{02HiQuxx.CH4,05Brxxxx.CH4,06BoReLo.CH4,08KaGaRo.CH4,09AlBaBo.CH4,09KaRoCa.CH4,
09ScKaGA.CH4,10VoMaPr.CH4,10CaWaKa.CH4,10WaKaLi.CH4,11LuMoKa.CH3D,11MoKaWa.CH4,
11NiThRe.CH4,11WaKaLi.CH4,12CaWaMo.CH4,12CaLeWa.CH4,12WaMoKa.CH4,13NiBoWe.CH4,13ReNiTy.CH4,
14UlBeAl.CH4,14VoMaPr.CH4,15BeKaCa.CH4,15BeLiCa.CH4, 16NiReTa.CH4,16ReNiCa.CH4,16AmLoPi.CH4} and there is strong motivation to continue working towards the visible region to aid the study of exoplanets.~\citep{Lupu:2016} Variational calculations from first principles were recently used in conjunction with an experimental line list~\citep{13CaLeWa.CH4} to assign a significant number of vibrational band centers in the icosad range ($6300${--}$7900\,$cm$^{-1}$).~\citep{16ReNiCa.CH4} This kind of analysis could prove extremely useful for more congested regions and its success depends on having a reliable potential energy surface (PES) to work with.

 The construction of highly accurate PESs for small polyatomic molecules has seen remarkable progress in recent years. It is now possible to compute vibrational energy levels within ``spectroscopic accuracy'' (better than $\pm 1{\,}$cm$^{-1}$) using a purely \textit{ab initio} PES.~\citep{03PoCsSh.H2O,02Scxxxx.CH4,YaYuRi11.H2CS,13MaKoXX.H2O2,15OwYuYa.CH3Cl,15OwYuYa.SiH4} To do so requires the use of a one-particle basis set near the complete basis set (CBS) limit, and the consideration of additional, higher-level (HL) contributions to recover more of the electron correlation energy.~\cite{QQC:2008,Peterson12} Although computationally demanding, these can be routinely calculated with most quantum chemistry codes.
 
 A number of accurate PESs for CH$_4$ have been reported in the literature.~\citep{95LeMaTa.CH4,98MaQuxx.CH4,04MaQuxx.CH4,99WaSixx.CH4,01ScPaxx.CH4,02Scxxxx.CH4,06OyYaTa.CH4,09WaScSh.CH4,11NiReTy.CH4,13YuTeBa.CH4,14YuTexx.CH4,14WaCaxx.CH4,15MaHeDa.CH4} These include purely \textit{ab initio} surfaces,~\citep{95LeMaTa.CH4,01ScPaxx.CH4,02Scxxxx.CH4,06OyYaTa.CH4,09WaScSh.CH4,15MaHeDa.CH4} and those which are based on \textit{ab initio} calculations but have subsequently been refined to experiment.~\citep{98MaQuxx.CH4,04MaQuxx.CH4,99WaSixx.CH4,11NiReTy.CH4,13YuTeBa.CH4,14YuTexx.CH4,14WaCaxx.CH4} The most rigorous \textit{ab initio} treatment to date was by \citet{02Scxxxx.CH4} who accounted for several HL contributions. Corrections to the full configuration interaction (CI) limit, core-valence (CV) electron correlation, scalar relativistic (SR) effects, the Lamb shift, the diagonal Born-Oppenheimer correction (DBOC), non-adiabatic corrections, as well as extrapolation of the basis set to the CBS limit, were all treated at some level. Whilst low-lying states of $^{12}$CH$_4$ were reproduced with sub-wavenumber accuracy, the description of the stretching fundamentals, $\nu_1$ and $\nu_3$, were relatively poor in comparison and the errors in vibrational energies gradually increased after $3000\,$cm$^{-1}$.

 As part of the ExoMol project~\cite{ExoMol2012,ExoMol2016} a comprehensive methane line list, 10to10,~\citep{14YuTexx.CH4} was produced by two of the authors. This line list represented a significant step forward in the variational treatment of five-atom molecules, and 10to10 has facilitated the detection of CH$_4$ in brown dwarfs,~\citep{14YuTexx.CH4} T dwarfs,~\citep{15CaLuYu.CH4} and the hot Jupiter exoplanet HD 189733b.~\citep{14YuTeBa.CH4} Since its construction a number of high resolution spectroscopic measurements on methane above the tetradecad region (above $6300\,$cm$^{-1}$) have been reported.~\citep{14VoMaPr.CH4,14UlBeAl.CH4,15BeKaCa.CH4,15BeLiCa.CH4,16NiReTa.CH4,16ReNiCa.CH4} There have also been key developments~\citep{15YaYu.ADF} in our nuclear motion code TROVE~\citep{TROVE2007} which considerably improves basis set convergence; a major bottleneck in the past. Given the demand for comprehensive methane data at higher energies and the knowledge we have acquired from the 10to10 line list, it seems natural to begin working on a more extensive and accurate treatment of CH$_4$.

 In this work we present a state-of-the-art \textit{ab initio} PES for methane. After fitting the \textit{ab initio} data with a symmetrized analytic representation, the PES is evaluated with variational calculations of pure rotational and $J\!=\!0$ energy levels. To ensure a reliable assessment, fully converged vibrational term values are obtained by means of a complete vibrational basis set (CVBS) extrapolation.~\cite{OvThYu08.PH3}

 The paper is structured as follows: In Sec.~\ref{sec:PES} the electronic structure calculations and analytic representation of the PES are presented. The variational nuclear motion computations used to validate the PES are described in Sec~\ref{sec:variational}. In Sec.~\ref{sec:results}, vibrational $J\!=\!0$ energy levels for $^{12}$CH$_4$, the equilibrium C{--}H bond length, and pure rotational energies up to $J=10$ are calculated and compared with available experimental results. We offer concluding remarks in Sec.~\ref{sec:conc}.
 
\section{Potential Energy Surface}
\label{sec:PES} 

\subsection{Electronic structure calculations}

 The approach employed for the electronic structure calculations is almost identical to our previous work on SiH$_4$.~\citep{15OwYuYa.SiH4} The aim is to generate a PES which has the `correct' shape and computing tightly converged energies with respect to basis set size for the HL corrections is not as important. The levels of theory and basis sets have therefore been chosen to strike a balance between accuracy and computational cost.
 
 Utilizing focal-point analysis~\citep{Csaszar98} the total electronic energy is written as
\begin{equation}\label{eq:tot_en}
E_{\mathrm{tot}} = E_{\mathrm{CBS}}+\Delta E_{\mathrm{CV}}+\Delta E_{\mathrm{HO}}+\Delta E_{\mathrm{SR}}+\Delta E_{\mathrm{DBOC}} .
\end{equation}
The energy at the complete basis set (CBS) limit $E_{\mathrm{CBS}}$ was computed using the explicitly correlated F12 coupled cluster method CCSD(T)-F12b~(Ref.~\onlinecite{Adler07}) in conjunction with the F12-optimized correlation consistent polarized valence basis sets, cc-pVTZ-F12 and cc-pVQZ-F12.~\cite{Peterson08} The frozen core approximation was employed and calculations used the diagonal fixed amplitude ansatz 3C(FIX)~\cite{TenNo04} with a Slater geminal exponent value of $\beta=1.0\,a_0^{-1}$.~\cite{Hill09} For the auxiliary basis sets (ABS), the OptRI,~\cite{Yousaf08} cc-pV5Z/JKFIT~\cite{Weigend02} and aug-cc-pwCV5Z/MP2FIT~\cite{Hattig05} were used for the resolution of the identity (RI) basis and the two density fitting (DF) basis sets, respectively. Calculations were carried out with MOLPRO2012~\cite{Werner2012} unless stated otherwise.

 To extrapolate to the CBS limit we used the parameterized, two-point 
formula~\cite{Hill09}
\begin{equation}\label{eq:cbs_extrap}
E^{C}_{\mathrm{CBS}} = (E_{n+1} - E_{n})F^{C}_{n+1} + E_{n} .
\end{equation}
The coefficients $F^{C}_{n+1}$, which are specific to the CCSD-F12b and (T) components of the total CCSD(T)-F12b energy, had values of $F^{\mathrm{CCSD-F12b}}=1.363388$ and $F^{\mathrm{(T)}}=1.769474$.~\citep{Hill09} No extrapolation was applied to the Hartree-Fock (HF) energy, rather the HF+CABS (complementary auxiliary basis set) singles correction~\cite{Adler07} calculated in the larger basis set was used.

 The contribution from core-valence (CV) electron correlation $\Delta E_{\mathrm{CV}}$ was computed at the CCSD(T)-F12b level of theory with the F12-optimized correlation consistent core-valence basis set cc-pCVTZ-F12.~\cite{Hill10} Calculations employed the same ansatz and ABS as used for $E_{\mathrm{CBS}}$, however, the Slater geminal exponent was changed to $\beta=1.4\,a_0^{-1}$. 
 
 Higher-order (HO) correlation effects were accounted for using the hierarchy of coupled cluster methods such that $\Delta E_{\mathrm{HO}} = \Delta E_{\mathrm{T}} + \Delta E_{\mathrm{(Q)}}$. Here, the full triples contribution is $\Delta E_{\mathrm{T}} = \left[E_{\mathrm{CCSDT}}-E_{\mathrm{CCSD(T)}}\right]$, and the perturbative quadruples contribution is $\Delta E_{\mathrm{(Q)}} = \left[E_{\mathrm{CCSDT(Q)}}-E_{\mathrm{CCSDT}}\right]$. Calculations were performed in the frozen core approximation at the CCSD(T), CCSDT, and CCSDT(Q) levels of theory using the general coupled cluster approach~\cite{Kallay05,Kallay08} as implemented in the MRCC code~\cite{mrcc} interfaced to CFOUR.~\cite{cfour} The correlation consistent triple zeta basis set, cc-pVTZ,~\cite{Dunning89} was utilized for the full triples contribution,whilst the perturbative quadruples employed the double zeta basis set, cc-pVDZ.
 
 The scalar relativistic (SR) correction $\Delta E_{\mathrm{SR}}$ was calculated with the second-order Douglas-Kroll-Hess approach~\cite{dk1,dk2} at the CCSD(T)/cc-pVQZ-DK~\cite{dk_basis} level of theory in the frozen core approximation. For light, closed-shell molecules the spin-orbit interaction can be neglected in spectroscopic calculations.~\cite{Tarczay01}
 
 The diagonal Born-Oppenheimer correction (DBOC) $\Delta E_{\mathrm{DBOC}}$ was computed with all electrons correlated using the CCSD method~\cite{Gauss06} as implemented in CFOUR with the aug-cc-pCVDZ basis set. The DBOC has a noticeable effect on vibrational term values of methane~\citep{02Scxxxx.CH4} but because it is mass dependent its inclusion means the PES is only applicable for $^{12}$CH$_4$.
 
 All terms in Eq.~\eqref{eq:tot_en} were calculated on a grid of $97{\,}721$ geometries with energies up to $h c \cdot 50{\,}000{\,}$cm$^{-1}$ ($h$ is the Planck constant and $c$ is the speed of light). The global grid was built in terms of nine internal coordinates; four C{--}H bond lengths $r_1$, $r_2$, $r_3$, $r_4$, and five $\angle(\mathrm{H}_j${--}C{--}$\mathrm{H}_k)$ interbond angles $\alpha_{12}$, $\alpha_{13}$, $\alpha_{14}$, $\alpha_{23}$, and $\alpha_{24}$, where $j$ and $k$ label the respective hydrogen atoms. The C{--}H stretch distances ranged from $0.71\leq r_i \leq 2.60{\,}\mathrm{\AA}$ for $i=1,2,3,4$ whilst bending angles varied from $40\leq \alpha_{jk} \leq 140^{\circ}$ where $jk=12,13,14,23,24$.

 Although it is computationally demanding to calculate the HL corrections at every grid point, it is actually time-effective given the system size, levels of theory and basis sets used. Timing data is shown in Table~\ref{tab:timing} and we see it takes just over $15$ minutes to compute all the contributions in Eq.~\eqref{eq:tot_en} at the equilibrium geometry. Naturally this time will increase as we stretch and bend the molecule due to slower energy convergence, with calculations needing at most $2${--}$3$ times longer for highly distorted geometries. 
 
 Alternatively, one can compute each HL correction on a reduced grid, fit a suitable analytic representation to the data and then interpolate to other points on the global grid (see Refs.~\onlinecite{YaYuRi11.H2CS,15OwYuYa.CH3Cl} for examples of this strategy). For more demanding systems this approach can significantly reduce computational time, however, obtaining an adequate description of each HL correction requires careful consideration and may not be straightforward. These issues are avoided in our present approach.
 
\begin{table}
\tabcolsep=0.35cm
\caption{\label{tab:timing}Wall clock times (seconds) for the different contributions to the potential energy surface. Calculations were performed on a single core of an Intel Xeon E5-2690 v2 $3.0\,$GHz processor. Timings shown have been averaged over 10 runs for one point at the equilibrium geometry.}
\begin{center}
\begin{tabular}{c c c}
\hline\hline
Contribution & No. of calculations required per point & Time\\
\hline
$E_{\mathrm{CBS}}$ & 2 & 296\\[-1.5mm]
$\Delta E_{\mathrm{CV}}$ & 2 & 107\\[-1.5mm]
$\Delta E_{\mathrm{HO}}$ & 3 & 234\\[-1.5mm]
$\Delta E_{\mathrm{SR}}$ & 2 & 189\\[-1.5mm]
$\Delta E_{\mathrm{DBOC}}$ & 1 & 87\\
\hline
$E_{\mathrm{tot}}$ & 10 & 913\\
\hline\hline
\end{tabular}
\end{center}
\end{table}
 
\subsection{Analytic representation}

 The XY$_4$ symmetrized analytic representation employed for the present study has previously been used for methane~\cite{13YuTeBa.CH4,14YuTexx.CH4} and silane.~\cite{15OwYuYa.SiH4} Morse oscillator functions describe the stretch coordinates,
\begin{equation}\label{eq:stretch}
\xi_i=1-\exp\left(-a(r_i - r_{\mathrm{ref}})\right){\,};\hspace{2mm}i=1,2,3,4 ,
\end{equation}
where $a=1.845{\,}\mathrm{\AA}^{-1}$ and the reference equilibrium structural parameter $r_{\mathrm{ref}}=1.08594{\,}\mathrm{\AA}$ (value discussed in Sec.~\ref{sec:eq_rotational}). For the angular terms we use symmetrized combinations of interbond angles,
\begin{equation}\label{eq:ang1}
\xi_5 = 
\frac{1}{\sqrt{12}}\left(2\alpha_{12}-\alpha_{13}-\alpha_{14}-\alpha_{23}-\alpha_{24}+2\alpha_{34}\right),
\end{equation}
\begin{equation}\label{eq:ang2}
\xi_6 = \frac{1}{2}\left(\alpha_{13}-\alpha_{14}-\alpha_{23}+\alpha_{24}\right),
\end{equation}
\begin{equation}\label{eq:ang3}
\xi_7 = \frac{1}{\sqrt{2}}\left(\alpha_{24}-\alpha_{13}\right),
\end{equation}
\begin{equation}\label{eq:ang4}
\xi_8 = \frac{1}{\sqrt{2}}\left(\alpha_{23}-\alpha_{14}\right),
\end{equation}
\begin{equation}\label{eq:ang5}
\xi_9 = \frac{1}{\sqrt{2}}\left(\alpha_{34}-\alpha_{12}\right).
\end{equation}

 The potential function,
\begin{equation}\label{eq:pot_f}
V(\xi_{1},\xi_{2},\xi_{3},\xi_{4},\xi_{5},\xi_{6},\xi_{7},\xi_{8},\xi_{9})={\sum_{ijk\ldots}}{\,}\mathrm{f}_{ijk\ldots}V_{ijk\ldots} ,
\end{equation}
which has maximum expansion order $i+j+k+l+m+n+p+q+r=6$, is composed of the terms
\begin{equation}\label{eq:pot_term}
V_{ijk\ldots}=\lbrace\xi_{1}^{\,i}\xi_{2}^{\,j}\xi_{3}^{\,k}\xi_{4}^{\,l}\xi_{5}^{\,m}\xi_{6}^{\,n}\xi_{7}^{\,p}\xi_{8}^{\,q}\xi_{9}^{\,r}\rbrace^{\bm{T}_{\mathrm{d}}\mathrm{(M)}} ,
\end{equation}
where $V_{ijk\ldots}$ are symmetrized combinations of different permutations of the coordinates $\xi_{i}$, and transform according to the $A_1$ representation of the $\bm{T}_{\mathrm{d}}\mathrm{(M)}$ molecular symmetry group.~\cite{MolSym_BuJe98} The terms in Eq.~\eqref{eq:pot_term} are found by solving an over-determined system of linear equations in terms of the nine coordinates given above. In total there are 287 symmetrically unique terms up to sixth order, of which only 110 were employed for the final PES. 

 A least-squares fitting to the \textit{ab initio} data was used to determine the expansion parameters $\mathrm{f}_{ijk\ldots}$. Weight factors of the form suggested by \citet{Schwenke97}
\begin{equation}\label{eq:weights}
w_i=\left(\frac{\tanh\left[-0.0006\times(\tilde{E}_i - 15{\,}000)\right]+1.002002002}{2.002002002}\right)\times\frac{1}{N\tilde{E}_i^{(w)}} ,
\end{equation}
were utilized in the fit. Here, $\tilde{E}_i^{(w)}=\max(\tilde{E}_i, 10{\,}000)$, where $\tilde{E}_i$ is the potential energy at the $i$th geometry above equilibrium and the normalization constant $N=0.0001$ (all values in cm$^{-1}$). In our fitting, energies below $15{\,}000{\,}$cm$^{-1}$ are favoured by the weighting scheme. To further improve the description at lower energies and reduce the weights of outliers we employed Watson's robust fitting scheme.~\cite{Watson03} The final PES was fitted with a weighted root-mean-square (rms) error of $1.08{\,}$cm$^{-1}$ for energies up to $h c \cdot 50{\,}000{\,}$cm$^{-1}$ and required 112 expansion parameters ($110+r_{\mathrm{ref}}+a$).

 For geometries where $r_i\geq 1.80{\,}\mathrm{\AA}$ for $i=1,2,3,4$, the respective weights were dropped by several orders of magnitude. At larger stretch distances a T1 diagnostic value $>0.02$ indicates that the coupled cluster method has become unreliable.~\cite{T1_Lee89} Energies are not wholly accurate at these points but they are still useful; their inclusion ensures the PES maintains a reasonable shape towards dissociation. In subsequent calculations we refer to this PES as CBS-F12$^{\,\mathrm{HL}}$. The CBS-F12$^{\,\mathrm{HL}}$ expansion parameter set is provided in the supplementary material along with a FORTRAN routine to construct the PES.~\cite{EPAPSCH4}

\section{Variational calculations}
\label{sec:variational}

 The general methodology of TROVE is well documented~\cite{TROVE2007,YuBaYa09.NH3,15YaYu.ADF} and calculations on methane have previously been reported.~\cite{13YuTeBa.CH4,14YuTexx.CH4} We therefore summarize only the key aspects relevant for this work.
 
 The rovibrational Hamiltonian was represented as a power series expansion around the equilibrium geometry in terms of the nine coordinates introduced in Eqs.~\eqref{eq:stretch} to \eqref{eq:ang5}. However, for the kinetic energy operator linear displacement variables $(r_i - r_{\mathrm{ref}})$ were used for the stretching coordinates. The Hamiltonian was constructed numerically using an automatic differentiation method~\cite{15YaYu.ADF} with the kinetic and potential energy operators truncated at 6th and 8th order, respectively. A discussion of the associated errors of such a scheme can be found in Refs.~\onlinecite{TROVE2007,15YaYu.ADF}. Atomic mass values were used throughout.
 
\begin{figure}
\includegraphics{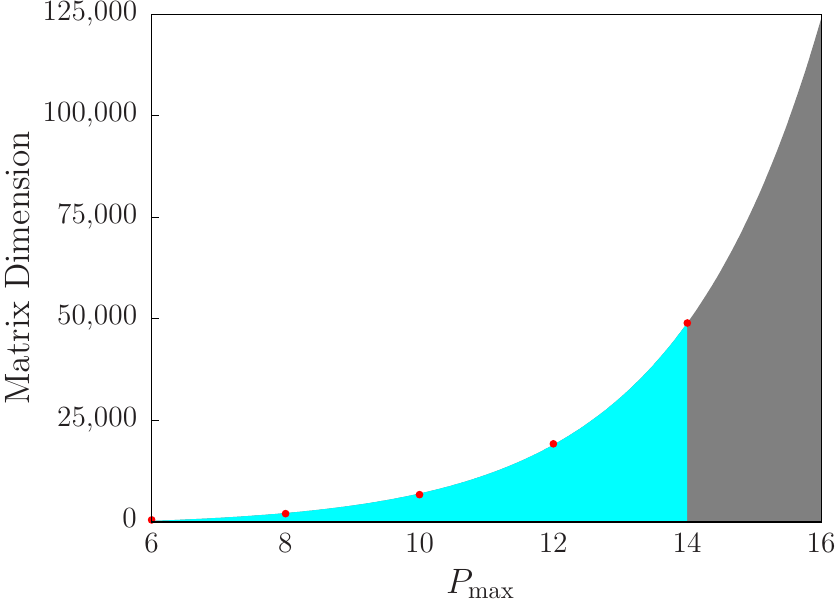}
\caption{\label{fig:dimension}Size of the $J\!=\!0$ Hamiltonian matrix with respect to the polyad truncation number $P_{\mathrm{max}}$. Calculations have not been possible above $P_{\mathrm{max}}=14$.}
\end{figure}

 A multi-step contraction scheme was employed to construct the vibrational basis set, the size of which is controlled by the polyad number,
\begin{equation}\label{eq:polyad}
P = 2(n_1+n_2+n_3+n_4)+n_5+n_6+n_7+n_8+n_9 \leq P_{\mathrm{max}} ,
\end{equation}
and this does not exceed a predefined maximum value $P_{\mathrm{max}}$. As shown in Fig.~\ref{fig:dimension}, the size of the Hamiltonian matrix grows exponentially with respect to $P_{\mathrm{max}}$ and calculations above $P_{\mathrm{max}}=14$ have not been possible with the resources available to us. Here the quantum numbers $n_k$ for $k=1,\ldots,9$ relate to primitive basis functions $\phi_{n_k}$, which are obtained by solving a one-dimensional Schr\"{o}dinger equation for each $k$th vibrational mode using the Numerov-Cooley method.~\cite{Numerov1924,Cooley1961} Multiplication with symmetrized rigid-rotor eigenfunctions $\ket{J,\Gamma_{\mathrm{rot}},n}$ gives the final basis set for use in $J>0$ calculations. The label $\Gamma_{\mathrm{rot}}$ is the rotational symmetry and $n$ is a multiplicity index used to count states within a given $J$ (see \citet{06BoReLo.CH4}).

 In TROVE the eigenvalues and corresponding eigenvectors are assigned with quantum numbers based on the contribution of the basis functions $\phi_{n_k}$. To be of spectroscopic use it is necessary to map these to the normal mode quantum numbers $\mathrm{v}_k$ commonly used. For CH$_4$, vibrational states are labelled as $\mathrm{v_1}\nu_1+\mathrm{v_2}\nu_2^{L_2}+\mathrm{v_3}\nu_3^{L_3}+\mathrm{v_4}\nu_4^{L_4}$ where $\mathrm{v}_i$ counts the level of excitation. The additional quantum numbers $L_i$ are the absolute values of the vibrational angular momentum quantum numbers $\ell_i$, which are needed to resolve the degeneracy of their respective modes (see \citet{14YuTexx.CH4} for further details). The non-degenerate symmetric stretching mode $\nu_{1}$ ($2916.48{\,}$cm$^{-1}$) is of $A_{1}$ symmetry. The doubly degenerate asymmetric bending mode $\nu_{2}$ ($1533.33{\,}$cm$^{-1}$) has $E$ symmetry. Whilst of $F_2$ symmetry are the triply degenerate modes; the asymmetric stretching mode $\nu_{3}$ ($3019.49{\,}$cm$^{-1}$), and the asymmetric bending mode $\nu_{4}$ ($1310.76{\,}$cm$^{-1}$). The values in parentheses are the experimentally determined values.~\citep{09AlBaBo.CH4}

\section{Results}
\label{sec:results}
 
\subsection{Vibrational $J\!=\!0$ energy levels}
\label{sec:J0_energies}

 A reliable assessment of the CBS-F12$^{\,\mathrm{HL}}$ PES is only possible with converged vibrational term values. Calculations with $P_{\mathrm{max}}=14$ are sufficient for converging low-lying states but this gradually deteriorates as we go up in energy. A way of overcoming this problem is to employ a complete vibrational basis set (CVBS) extrapolation.~\cite{OvThYu08.PH3} Similar to basis set extrapolation techniques of electronic structure theory,~\cite{Petersson88,Petersson91} the same approach can be applied to TROVE calculations with respect to $P_{\mathrm{max}}$. We use the exponential decay expression,
\begin{equation}
 E_i(P_{\mathrm{max}}) = E_i^{\mathrm{CVBS}}+A_i\exp(-\lambda_i P_{\mathrm{max}}) ,
\end{equation}
where $E_i$ is the energy of the $i$th level, $E_i^{\mathrm{CVBS}}$ is the corresponding energy at the CVBS limit, $A_i$ is a fitting parameter, $\lambda_i$ is determined from
\begin{equation}
\lambda_i=-\frac{1}{2}\ln\left(\frac{E_i(P_{\mathrm{max}}+2)-E_i(P_{\mathrm{max}})}{E_i(P_{\mathrm{max}})-E_i(P_{\mathrm{max}}-2)}\right) ,
\end{equation}
and the values of $P_{\mathrm{max}}=\lbrace 10,12,14 \rbrace$.

 Briefly commenting on the accuracy of the CVBS extrapolation itself, similar to electronic structure theory the use of larger basis sets is always preferable for the extrapolation. Highly excited modes benefit the most as convergence is much slower, however, at higher energies the increased density of states makes it harder to consistently identify and match energy levels for different values of $P_{\mathrm{max}}$. To ensure a reliable extrapolation we have also found that $\lambda_i\geq0.5$.

 In the following comparisons we have collected, to the best of our knowledge, all $J\!=\!0$ energies that have been accurately determined from experiment (see \citet{12MaQuxx.CH4} for a discussion of the experimental uncertainties associated with methane spectra). Although very minor discrepancies occasionally occur between different studies, the majority of vibrational term values up to the tetradecad region (up to $6300\,$cm$^{-1}$) are fairly well established. Progress is being made in the icosad range ($6300${--}$7900\,$cm$^{-1}$) and a large number of levels have recently been assigned~\citep{16NiReTa.CH4,16ReNiCa.CH4} using the WKLMC line list.~\citep{13CaLeWa.CH4} At even higher energies several vibrational band centers have been measured and assigned by means of an assignment of their P(1) transitions up to about $11\,300\,$cm$^{-1}$.~\citep{14UlBeAl.CH4}
 
 Computed vibrational energy levels for $^{12}$CH$_4$ up to the tetradecad region are listed in Table~\ref{tab:j0_12ch4_1}. The four fundamentals are reproduced with a rms error of $0.70\,$cm$^{-1}$ and a mean-absolute-deviation (mad) of $0.64\,$cm$^{-1}$. Around $70\%$ of the $89$ term values are calculated within spectroscopic accuracy (better than $\pm 1\,$cm$^{-1}$) and this does not include the $4\nu_4$ levels computed at $P_{\mathrm{max}}=14$, which are not fully converged. 
 
\LTcapwidth=\textwidth
\begin{longtable}{@{\extracolsep{0.2cm}} l c c c r c}
\caption{\label{tab:j0_12ch4_1}Comparison of calculated and experimental $J\!=\!0$ vibrational term values (in cm$^{-1}$) up to the tetradecad region for $^{12}$CH$_4$. The zero-point energy was computed to be $9708.846\,$cm$^{-1}$ at the CVBS limit.}\\ \hline\hline
Mode & Sym. &  Experiment & Calculated & Obs$-$calc & Ref.\\ \hline
\endfirsthead
\caption{(\textit{Continued})}\\ \hline 
Mode & Sym. &  Experiment & Calculated & Obs$-$calc & Ref.\\ \hline
\endhead
$\nu_4^{\it 1}$                             & $F_2$ & 1310.76&          1310.24&  0.52& \onlinecite{09AlBaBo.CH4}\\[-1.5mm]
$\nu_2^{\it 1}$                             & $E$   & 1533.33&          1533.04&  0.29& \onlinecite{09AlBaBo.CH4}\\[-1.5mm]
$2\nu_4^{\it 0}$                            & $A_1$ & 2587.04&          2585.74&  1.30& \onlinecite{09AlBaBo.CH4}\\[-1.5mm]
$2\nu_4^{\it 2}$                            & $F_2$ & 2614.26&          2613.04&  1.22& \onlinecite{09AlBaBo.CH4}\\[-1.5mm]
$2\nu_4^{\it 2}$                            & $E$   & 2624.62&          2624.08&  0.54& \onlinecite{09AlBaBo.CH4}\\[-1.5mm]
$\nu_2^{\it 1}+\nu_4^{\it 1}$               & $F_2$ & 2830.32&          2829.71&  0.61& \onlinecite{09AlBaBo.CH4}\\[-1.5mm]
$\nu_2^{\it 1}+\nu_4^{\it 1}$               & $F_1$ & 2846.07&          2845.44&  0.63& \onlinecite{09AlBaBo.CH4}\\[-1.5mm]
$\nu_1$                                     & $A_1$ & 2916.48&          2917.16& -0.68& \onlinecite{09AlBaBo.CH4}\\[-1.5mm]
$\nu_3^{\it 1}$                             & $F_2$ & 3019.49&          3020.57& -1.08& \onlinecite{09AlBaBo.CH4}\\[-1.5mm]
$2\nu_2^{\it 0}$                            & $A_1$ & 3063.65&          3063.04&  0.61& \onlinecite{09AlBaBo.CH4}\\[-1.5mm]
$2\nu_2^{\it 2}$                            & $E$   & 3065.14&          3064.53&  0.61& \onlinecite{09AlBaBo.CH4}\\[-1.5mm]
$3\nu_4^{\it 1}$                            & $F_2$ & 3870.49&          3869.18&  1.31& \onlinecite{09AlBaBo.CH4}\\[-1.5mm]
$3\nu_4^{\it 1}$                            & $A_1$ & 3909.20&          3907.11&  2.09& \onlinecite{09AlBaBo.CH4}\\[-1.5mm]
$3\nu_4^{\it 3}$                            & $F_1$ & 3920.50&          3919.01&  1.49& \onlinecite{13NiBoWe.CH4}\\[-1.5mm]
$3\nu_4^{\it 3}$                            & $F_2$ & 3930.92&          3930.00&  0.92& \onlinecite{09AlBaBo.CH4}\\[-1.5mm]
$\nu_2^{\it 1}+2\nu_4^{\it 0}$              & $E$   & 4101.39&          4100.52&  0.87& \onlinecite{09AlBaBo.CH4}\\[-1.5mm]
$\nu_2^{\it 1}+2\nu_4^{\it 2}$              & $F_1$ & 4128.77&          4127.77&  1.00& \onlinecite{13NiBoWe.CH4}\\[-1.5mm]
$\nu_2^{\it 1}+2\nu_4^{\it 2}$              & $A_1$ & 4132.88&          4132.21&  0.67& \onlinecite{13NiBoWe.CH4}\\[-1.5mm]
$\nu_2^{\it 1}+2\nu_4^{\it 2}$              & $F_2$ & 4142.86&          4142.03&  0.83& \onlinecite{13NiBoWe.CH4}\\[-1.5mm]
$\nu_2^{\it 1}+2\nu_4^{\it 2}$              & $E$   & 4151.20&          4150.62&  0.58& \onlinecite{13NiBoWe.CH4}\\[-1.5mm]
$\nu_2^{\it 1}+2\nu_4^{\it 2}$              & $A_2$ & 4161.84&          4161.00&  0.84& \onlinecite{13NiBoWe.CH4}\\[-1.5mm]
$\nu_1+\nu_4^{\it 1}$                       & $F_2$ & 4223.46&          4223.62& -0.16& \onlinecite{09AlBaBo.CH4}\\[-1.5mm]
$\nu_3^{\it 1}+\nu_4^{\it 1}$               & $F_2$ & 4319.21&          4319.37& -0.16& \onlinecite{09AlBaBo.CH4}\\[-1.5mm]
$\nu_3^{\it 1}+\nu_4^{\it 1}$               & $E$   & 4322.18&          4323.38& -1.20& \onlinecite{09AlBaBo.CH4}\\[-1.5mm]
$\nu_3^{\it 1}+\nu_4^{\it 1}$               & $F_1$ & 4322.58&          4323.53& -0.95& \onlinecite{13NiBoWe.CH4}\\[-1.5mm]
$\nu_3^{\it 1}+\nu_4^{\it 1}$               & $A_1$ & 4322.72&          4323.01& -0.29& \onlinecite{13NiBoWe.CH4}\\[-1.5mm]
$2\nu_2^{\it 0}+\nu_4^{\it 1}$              & $F_2$ & 4348.72&          4348.07&  0.65& \onlinecite{09AlBaBo.CH4}\\[-1.5mm]
$2\nu_2^{\it 2}+\nu_4^{\it 1}$              & $F_1$ & 4363.62&          4362.86&  0.76& \onlinecite{13NiBoWe.CH4}\\[-1.5mm]
$2\nu_2^{\it 2}+\nu_4^{\it 1}$              & $F_2$ & 4378.94&          4378.30&  0.64& \onlinecite{13NiBoWe.CH4}\\[-1.5mm]
$\nu_1+\nu_2^{\it 1}$                       & $E$   & 4435.13&          4435.25& -0.12& \onlinecite{13NiBoWe.CH4}\\[-1.5mm]
$\nu_2^{\it 1}+\nu_3^{\it 1}$               & $F_1$ & 4537.55&          4538.13& -0.58& \onlinecite{09AlBaBo.CH4}\\[-1.5mm]
$\nu_2^{\it 1}+\nu_3^{\it 1}$               & $F_2$ & 4543.76&          4544.36& -0.60& \onlinecite{09AlBaBo.CH4}\\[-1.5mm]
$3\nu_2^{\it 1}$                            & $E$   & 4592.03&          4591.08&  0.95& \onlinecite{09AlBaBo.CH4}\\[-1.5mm]
$3\nu_2^{\it 3}$                            & $A_2$ & 4595.28&          4594.40&  0.88& \onlinecite{13NiBoWe.CH4}\\[-1.5mm]
$3\nu_2^{\it 3}$                            & $A_1$ & 4595.52&          4594.49&  1.03& \onlinecite{13NiBoWe.CH4}\\[-1.5mm]
$4\nu_4^{\it 0}$                            & $A_1$ & 5121.77& 5121.51$^{\,a}$ &  0.26& \onlinecite{16AmLoPi.CH4}\\[-1.5mm]
$4\nu_4^{\it 2}$                            & $F_2$ & 5143.36& 5143.07$^{\,a}$ &  0.29& \onlinecite{13NiBoWe.CH4}\\[-1.5mm]
$4\nu_4^{\it 2}$                            & $E$   & 5167.20& 5167.15$^{\,a}$ &  0.05& \onlinecite{13NiBoWe.CH4}\\[-1.5mm]
$4\nu_4^{\it 4}$                            & $F_2$ & 5210.74& 5209.06$^{\,a}$ &  1.68& \onlinecite{13NiBoWe.CH4}\\[-1.5mm]
$4\nu_4^{\it 4}$                            & $E$   & 5228.74& 5227.45$^{\,a}$ &  1.29& \onlinecite{13NiBoWe.CH4}\\[-1.5mm]
$4\nu_4^{\it 4}$                            & $F_1$ & 5230.59& 5229.46$^{\,a}$ &  1.13& \onlinecite{16AmLoPi.CH4}\\[-1.5mm]
$4\nu_4^{\it 4}$                            & $A_1$ & 5240.46& 5239.76$^{\,a}$ &  0.70& \onlinecite{16AmLoPi.CH4}\\[-1.5mm]
$\nu_2^{\it 1}+3\nu_4^{\it 1}$              & $F_2$ & 5370.48&          5369.79&  0.69& \onlinecite{16AmLoPi.CH4}\\[-1.5mm]
$\nu_2^{\it 1}+3\nu_4^{\it 1}$              & $F_1$ & 5389.74&          5388.96&  0.78& \onlinecite{16AmLoPi.CH4}\\[-1.5mm]
$\nu_2^{\it 1}+3\nu_4^{\it 1}$              & $E$   & 5424.80&          5423.39&  1.41& \onlinecite{16AmLoPi.CH4}\\[-1.5mm]
$\nu_2^{\it 1}+3\nu_4^{\it 3}$              & $F_2$ & 5429.86&          5428.85&  1.01& \onlinecite{16AmLoPi.CH4}\\[-1.5mm]
$\nu_2^{\it 1}+3\nu_4^{\it 3}$              & $F_1$ & 5437.28&          5436.38&  0.90& \onlinecite{16AmLoPi.CH4}\\[-1.5mm]
$\nu_2^{\it 1}+3\nu_4^{\it 3}$              & $F_2$ & 5444.80&          5444.07&  0.73& \onlinecite{13NiBoWe.CH4}\\[-1.5mm]
$\nu_2^{\it 1}+3\nu_4^{\it 3}$              & $F_1$ & 5462.91&          5461.86&  1.05& \onlinecite{16AmLoPi.CH4}\\[-1.5mm]
$\nu_1+2\nu_4^{\it 0}$                      & $A_1$ & 5492.90&          5492.32&  0.58& \onlinecite{16AmLoPi.CH4}\\[-1.5mm]
$\nu_3^{\it 1}+2\nu_4^{\it 0}$              & $F_2$ & 5587.97&          5587.97&  0.00& \onlinecite{13NiBoWe.CH4}\\[-1.5mm]
$\nu_3^{\it 1}+2\nu_4^{\it 2}$              & $A_1$ & 5604.47&          5604.51& -0.04& \onlinecite{13NiBoWe.CH4}\\[-1.5mm]
$2\nu_2^{\it 0}+2\nu_4^{\it 0}$             & $A_1$ & 5613.88&          5612.61&  1.27& \onlinecite{16AmLoPi.CH4}$^{\,b}$\\[-1.5mm]
$2\nu_2^{\it 2}+2\nu_4^{\it 0}$             & $E$   & 5614.58&          5613.15&  1.43& \onlinecite{16AmLoPi.CH4}\\[-1.5mm]
$\nu_3^{\it 1}+2\nu_4^{\it 2}$              & $F_1$ & 5615.37&          5615.75& -0.38& \onlinecite{16AmLoPi.CH4}\\[-1.5mm]
$\nu_3^{\it 1}+2\nu_4^{\it 2}$              & $F_2$ & 5616.02&          5615.46&  0.56& \onlinecite{16AmLoPi.CH4}\\[-1.5mm]
$\nu_3^{\it 1}+2\nu_4^{\it 2}$              & $E$   & 5618.23&          5618.85& -0.62& \onlinecite{16AmLoPi.CH4}\\[-1.5mm]
$\nu_3^{\it 1}+2\nu_4^{\it 2}$              & $F_1$ & 5626.10&          5626.96& -0.86& \onlinecite{16AmLoPi.CH4}\\[-1.5mm]
$\nu_3^{\it 1}+2\nu_4^{\it 2}$              & $F_2$ & 5627.35&          5628.29& -0.94& \onlinecite{16AmLoPi.CH4}\\[-1.5mm]
$2\nu_2^{\it 0}+2\nu_4^{\it 2}$             & $F_2$ & 5641.88&          5641.63&  0.25& \onlinecite{16AmLoPi.CH4}\\[-1.5mm]
$2\nu_2^{\it 2}+2\nu_4^{\it 2}$             & $E$   & 5654.47&          5653.58&  0.89& \onlinecite{16AmLoPi.CH4}\\[-1.5mm]
$2\nu_2^{\it 2}+2\nu_4^{\it 2}$             & $F_1$ & 5655.76&          5655.28&  0.48& \onlinecite{13NiBoWe.CH4}\\[-1.5mm]
$2\nu_2^{\it 2}+2\nu_4^{\it 2}$             & $A_2$ & 5664.08&          5663.38&  0.70& \onlinecite{16AmLoPi.CH4}\\[-1.5mm]
$2\nu_2^{\it 0}+2\nu_4^{\it 2}$             & $F_2$ & 5668.33&          5668.25&  0.08& \onlinecite{16AmLoPi.CH4}\\[-1.5mm]
$2\nu_2^{\it 2}+2\nu_4^{\it 2}$             & $A_1$ & 5681.26&          5681.25&  0.01& \onlinecite{16AmLoPi.CH4}\\[-1.5mm]
$2\nu_2^{\it 0}+2\nu_4^{\it 2}$             & $E$   & 5691.10&          5690.32&  0.78& \onlinecite{16AmLoPi.CH4}\\[-1.5mm]
$2\nu_1$                                    & $A_1$ & 5790.25&          5792.08& -1.83& \onlinecite{97MaBeSa.CH4}\\[-1.5mm]
$\nu_2^{\it 1}+\nu_3^{\it 1}+\nu_4^{\it 1}$ & $F_2$ & 5823.10&          5823.65& -0.55& \onlinecite{13NiBoWe.CH4}\\[-1.5mm]
$\nu_2^{\it 1}+\nu_3^{\it 1}+\nu_4^{\it 1}$ & $F_1$ & 5825.43&          5825.59& -0.16& \onlinecite{16AmLoPi.CH4}\\[-1.5mm]
$\nu_2^{\it 1}+\nu_3^{\it 1}+\nu_4^{\it 1}$ & $E$   & 5832.02&          5832.60& -0.58& \onlinecite{13NiBoWe.CH4}\\[-1.5mm]
$\nu_2^{\it 1}+\nu_3^{\it 1}+\nu_4^{\it 1}$ & $A_1$ & 5834.82&          5835.64& -0.82& \onlinecite{13NiBoWe.CH4}\\[-1.5mm]
$\nu_2^{\it 1}+\nu_3^{\it 1}+\nu_4^{\it 1}$ & $E$   & 5842.57&          5843.12& -0.55& \onlinecite{16AmLoPi.CH4}\\[-1.5mm]
$\nu_2^{\it 1}+\nu_3^{\it 1}+\nu_4^{\it 1}$ & $A_2$ & 5843.19&          5843.83& -0.64& \onlinecite{16AmLoPi.CH4}\\[-1.5mm]
$\nu_2^{\it 1}+\nu_3^{\it 1}+\nu_4^{\it 1}$ & $F_2$ & 5844.03&          5844.28& -0.25& \onlinecite{13NiBoWe.CH4}\\[-1.5mm]
$\nu_2^{\it 1}+\nu_3^{\it 1}+\nu_4^{\it 1}$ & $F_1$ & 5847.39&          5847.66& -0.27& \onlinecite{16AmLoPi.CH4}\\[-1.5mm]
$\nu_1+\nu_3^{\it 1}$                       & $F_2$ & 5861.49&          5861.90& -0.41& \onlinecite{13NiBoWe.CH4}\\[-1.5mm]
$3\nu_2^{\it 1}+\nu_4^{\it 1}$              & $F_2$ & 5867.52&          5868.09& -0.57& \onlinecite{16AmLoPi.CH4}\\[-1.5mm]
$3\nu_2^{\it 3}+\nu_4^{\it 1}$              & $F_1$ & 5879.80&          5878.97&  0.83& \onlinecite{16AmLoPi.CH4}\\[-1.5mm]
$3\nu_2^{\it 3}+\nu_4^{\it 1}$              & $F_2$ & 5894.34&          5893.51&  0.83& \onlinecite{16AmLoPi.CH4}\\[-1.5mm]
$3\nu_2^{\it 1}+\nu_4^{\it 1}$              & $F_1$ & 5908.74&          5908.52&  0.22& \onlinecite{16AmLoPi.CH4}\\[-1.5mm]
$\nu_1+2\nu_2^{\it 2}$                      & $E$   & 5952.44&          5952.00&  0.44& \onlinecite{13NiBoWe.CH4}\\[-1.5mm]
$2\nu_3^{\it 0}$                            & $A_1$ & 5968.15&          5969.12& -0.97& \onlinecite{98GeHeHi.CH4}\\[-1.5mm]
$2\nu_3^{\it 2}$                            & $F_2$ & 6004.62&          6006.54& -1.92& \onlinecite{13NiBoWe.CH4}\\[-1.5mm]
$2\nu_3^{\it 2}$                            & $E$   & 6043.82&          6046.12& -2.30& \onlinecite{13NiBoWe.CH4}\\[-1.5mm]
$2\nu_2^{\it 0}+\nu_3^{\it 1}$              & $F_2$ & 6054.61&          6054.74& -0.13& \onlinecite{13NiBoWe.CH4}\\[-1.5mm]
$2\nu_2^{\it 2}+\nu_3^{\it 1}$              & $F_1$ & 6060.62&          6060.67& -0.05& \onlinecite{13NiBoWe.CH4}\\[-1.5mm]
$2\nu_2^{\it 2}+\nu_3^{\it 1}$              & $F_2$ & 6065.59&          6065.48&  0.11& \onlinecite{13NiBoWe.CH4}\\[-1.5mm]
$4\nu_2^{\it 2}$                            & $E$   & 6118.95&          6117.21&  1.74& \onlinecite{16AmLoPi.CH4}\\[-1.5mm]
$4\nu_2^{\it 4}$                            & $E$   & 6124.12&          6122.77&  1.35& \onlinecite{16AmLoPi.CH4}\\
\hline\hline
\caption*{$^a$ $P_{\mathrm{max}}=14$ value. $^b$ Assigned as $\nu_3+2\nu_4$ in TROVE.\\}
\end{longtable}

 Six energy levels in the tetradecad region have not been included in Table~\ref{tab:j0_12ch4_1} because their experimental uncertainty could be as large as $5\,$cm$^{-1}$ (see \citet{13NiBoWe.CH4}). Instead they are listed in Table~\ref{tab:j0_six} alongside computed values from the CBS-F12$^{\,\mathrm{HL}}$ PES, the empirically refined PES of \citet{14WaCaxx.CH4} (denoted as WC), and the empirically adjusted PES of \citet{11NiReTy.CH4} (denoted as NRT). The three PESs show consistent agreement with each other, notably for the $\nu_1+2\nu_2^{\it 0}(A_1)$ and $4\nu_2^{\it 0}(A_1)$ levels where the residual errors, $\Delta E(\mathrm{obs}-\mathrm{calc})$, compared to \citet{13NiBoWe.CH4} are the largest. This would suggest that the effective Hamiltonian model used in \citet{13NiBoWe.CH4} and subsequently updated by \citet{16AmLoPi.CH4} may need further refinement in the tetradecad region.
 
\begin{table}
\tabcolsep=0.35cm
\caption{\label{tab:j0_six}Six $J\!=\!0$ vibrational term values (in cm$^{-1}$) in the tetradecad region which have a large experimental uncertainty (see text). Comparisons are given with the CBS-F12$^{\,\mathrm{HL}}$ PES (this work), the empirically refined PES of \citet{14WaCaxx.CH4} (denoted as WC), and the empirically adjusted PES of \citet{11NiReTy.CH4} (denoted as NRT).}
\begin{center}
\begin{tabular}{l c c c c c c}
\hline\hline
Mode & Sym. &  Experiment~\citep{13NiBoWe.CH4} & CBS-F12$^{\,\mathrm{HL}}$ & WC& NRT \\
\hline
$\nu_1+2\nu_4^{\it 2}$ & $F_2$ & 5519.88 & 5520.95 & 5522.32 & 5522.66 \\[-1.5mm]
$\nu_1+2\nu_4^{\it 2}$ & $E$   & 5536.23 & 5533.62 & 5534.54 & 5534.20 \\[-1.5mm]
$\nu_1+\nu_2^{\it 1}+\nu_4^{\it 1}$ & $F_2$ & 5728.58 & 5726.71 & 5727.50 & 5727.72 \\[-1.5mm]
$\nu_1+\nu_2^{\it 1}+\nu_4^{\it 1}$ & $F_1$ & 5745.90 & 5744.72 & 5745.78 & 5745.31 \\[-1.5mm]
$\nu_1+2\nu_2^{\it 0}$ & $A_1$ & 5945.81 & 5940.11 & 5939.90 & 5939.96 \\[-1.5mm]
$4\nu_2^{\it 0}$ & $A_1$ & 6122.13 & 6115.42 & 6116.74 & 6117.75 \\
\hline\hline
\end{tabular}
\end{center}
\end{table}

 For the icosad region and above, shown in Table~\ref{tab:j0_12ch4_2} and Table~\ref{tab:j0_12ch4_3}, spectroscopic accuracy is again achieved for around $70\%$ of the $134$ term values considered. Here we have separated the computed energies into two separate tables based on the accuracy of the corresponding values from experiment, which are predominantly from Refs.~\onlinecite{11NiThRe.CH4,16NiReTa.CH4,16ReNiCa.CH4}. The values in Table~\ref{tab:j0_12ch4_2} have an experimental accuracy of $0.0015\,$cm$^{-1}$ (the $\nu_2^{\it 1}+2\nu_3^{\it 2}$ level from \citet{02HiQuxx.CH4} has an uncertainty of $0.0010\,$cm$^{-1}$). In Table~\ref{tab:j0_12ch4_3}, energies have an accuracy of $0.1${--}$0.4\,$cm$^{-1}$, except for the vibrational band centers from \citet{14UlBeAl.CH4} which have a reported experimental uncertainty of around $0.001\,$cm$^{-1}$; a result of the direct method used. However, the $\nu_1+\nu_3^{\it 1}+\nu_4^{\it 1}(F_2)$ level from \citet{14UlBeAl.CH4} shows a discrepancy of $1.41\,$cm$^{-1}$ compared to the recent value published by \citet{16ReNiCa.CH4}.
 
\begin{figure}
\includegraphics{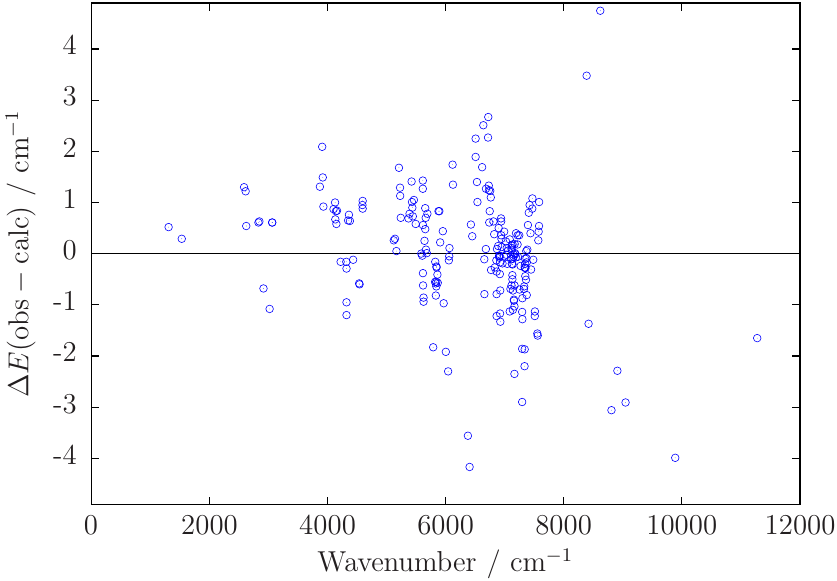}
\caption{\label{fig:res_vib}Residual errors $\Delta E(\mathrm{obs}-\mathrm{calc})$ for all computed term values of $^{12}$CH$_4$ (see Tables~\ref{tab:j0_12ch4_1}, \ref{tab:j0_12ch4_2} and \ref{tab:j0_12ch4_3}).}
\end{figure}

 Three term values from \citet{14UlBeAl.CH4} above $10\,000\,$cm$^{-1}$ could not be confidently identified in TROVE. The increased density of states and approximate TROVE labelling scheme can make it difficult to unambiguously discern certain levels. Regardless, from Table~\ref{tab:j0_12ch4_2} and Table~\ref{tab:j0_12ch4_3} it is evident that the CBS-F12$^{\,\mathrm{HL}}$ PES provides a reliable description at higher energies and there does not appear to be any significant deterioration in accuracy (see Fig.~\ref{fig:res_vib} for an overview of the residual errors for all term values). This will be important for investigating methane spectra up to the $14\,000\,$cm$^{-1}$ region, which is a key motivation for the present work.

\LTcapwidth=\textwidth
\begin{longtable}{@{\extracolsep{0.2cm}} l c c c r c}
\caption{\label{tab:j0_12ch4_2}Comparison of calculated and experimental $J\!=\!0$ vibrational term values (in cm$^{-1}$) for $^{12}$CH$_4$ in the icosad region (see text for a discussion of the experimental uncertainties). The zero-point energy was computed to be $9708.846\,$cm$^{-1}$ at the CVBS limit.}\\ \hline\hline
Mode & Sym. &  Experiment & Calculated & Obs$-$calc & Ref.\\ \hline
\endfirsthead
\caption{(\textit{Continued})}\\ \hline 
Mode & Sym. &  Experiment & Calculated & Obs$-$calc & Ref.\\ \hline
\endhead
$5\nu_4^{\it 1}$                              &  $F_2$ &  6450.06&        6449.72&  0.34&  \onlinecite{11NiThRe.CH4}\\[-1.5mm]
$5\nu_4^{\it 5}$                              &  $F_2$ &  6507.55&        6505.66&  1.89&  \onlinecite{11NiThRe.CH4}\\[-1.5mm]
$5\nu_4^{\it 5}$                              &  $F_2$ &  6539.18&        6538.17&  1.01&  \onlinecite{11NiThRe.CH4}\\[-1.5mm]
$\nu_2^{\it 1}+4\nu_4^{\it 2}$                &  $F_2$ &  6657.09&  6657.88$^{\,a}$& -0.79&  \onlinecite{16NiReTa.CH4}\\[-1.5mm]
$\nu_2^{\it 1}+4\nu_4^{\it 4}$                &  $F_2$ &  6717.99&        6715.72&  2.27&  \onlinecite{16ReNiCa.CH4}\\[-1.5mm]
$\nu_2^{\it 1}+4\nu_4^{\it 4}$                &  $F_2$ &  6733.11&        6731.87&  1.24&  \onlinecite{16ReNiCa.CH4}\\[-1.5mm]
$\nu_1+3\nu_4^{\it 1}$                        &  $F_2$ &  6769.19&        6769.51& -0.32&  \onlinecite{16ReNiCa.CH4}\\[-1.5mm]
$\nu_1+3\nu_4^{\it 3}$                        &  $F_2$ &  6833.19&        6833.46& -0.27&  \onlinecite{16ReNiCa.CH4}\\[-1.5mm]
$\nu_3^{\it 1}+3\nu_4^{\it 1}$                &  $F_2$ &  6858.71&        6858.84& -0.13&  \onlinecite{16ReNiCa.CH4}\\[-1.5mm]
$2\nu_2^{\it 0}+3\nu_4^{\it 1}$               &  $F_2$ &  6869.79&        6869.70&  0.09&  \onlinecite{16ReNiCa.CH4}\\[-1.5mm]
$\nu_3^{\it 1}+3\nu_4^{\it 3}$                &  $F_2$ &  6897.38&        6896.88&  0.50&  \onlinecite{16ReNiCa.CH4}\\[-1.5mm]
$\nu_3^{\it 1}+3\nu_4^{\it 1}$                &  $F_2$ &  6910.38&        6910.46& -0.08&  \onlinecite{16ReNiCa.CH4}\\[-1.5mm]
$\nu_3^{\it 1}+3\nu_4^{\it 1}$                &  $F_2$ &  6924.97&        6925.69& -0.72&  \onlinecite{16ReNiCa.CH4}\\[-1.5mm]
$2\nu_2^{\it 2}+3\nu_4^{\it 3}$               &  $F_2$ &  6940.05&        6939.69&  0.36&  \onlinecite{16NiReTa.CH4}\\[-1.5mm]
$2\nu_2^{\it 2}+3\nu_4^{\it 3}$               &  $F_2$ &  6992.58&        6992.15&  0.43&  \onlinecite{16ReNiCa.CH4}\\[-1.5mm]
$\nu_1+\nu_2^{\it 1}+2\nu_4^{\it 2}$          &  $F_2$ &  7035.18&        7035.07&  0.11&  \onlinecite{16ReNiCa.CH4}\\[-1.5mm]
$2\nu_1+\nu_4^{\it 1}$                        &  $F_2$ &  7085.64&        7086.77& -1.13&  \onlinecite{16ReNiCa.CH4}\\[-1.5mm]
$\nu_2^{\it 1}+\nu_3^{\it 1}+2\nu_4^{\it 0}$  &  $F_2$ &  7097.92&        7098.61& -0.69&  \onlinecite{16ReNiCa.CH4}$^{\,b}$\\[-1.5mm]
$\nu_2^{\it 1}+\nu_3^{\it 1}+2\nu_4^{\it 2}$  &  $F_2$ &  7116.39&        7117.01& -0.62&  \onlinecite{16ReNiCa.CH4}\\[-1.5mm]
$\nu_2^{\it 1}+\nu_3^{\it 1}+2\nu_4^{\it 2}$  &  $F_2$ &  7131.14&        7131.56& -0.42&  \onlinecite{16ReNiCa.CH4}\\[-1.5mm]
$\nu_1+\nu_3^{\it 1}+\nu_4^{\it 1}$           &  $F_2$ &  7158.13&        7159.05& -0.92&  \onlinecite{16ReNiCa.CH4}$^{\,c}$\\[-1.5mm]
$3\nu_2^{\it 3}+2\nu_4^{\it 2}$               &  $F_2$ &  7168.42&        7168.23&  0.19&  \onlinecite{16ReNiCa.CH4}\\[-1.5mm]
$\nu_1+2\nu_2^{\it 2}+\nu_4^{\it 1}$          &  $F_2$ &  7225.43&        7225.49& -0.06&  \onlinecite{16ReNiCa.CH4}\\[-1.5mm]
$2\nu_3^{\it 0}+\nu_4^{\it 1}$                &  $F_2$ &  7250.54&        7251.24& -0.70&  \onlinecite{16ReNiCa.CH4}\\[-1.5mm]
$\nu_1+2\nu_2^{\it 0}+\nu_4^{\it 1}$          &  $F_2$ &  7269.44&        7269.68& -0.24&  \onlinecite{16ReNiCa.CH4}\\[-1.5mm]
$2\nu_3^{\it 2}+\nu_4^{\it 1}$                &  $F_2$ &  7299.44&        7300.72& -1.28&  \onlinecite{16ReNiCa.CH4}\\[-1.5mm]
$2\nu_2^{\it 0}+\nu_3^{\it 1}+\nu_4^{\it 1}$  &  $F_2$ &  7331.05&        7331.69& -0.64&  \onlinecite{16ReNiCa.CH4}\\[-1.5mm]
$2\nu_2^{\it 2}+\nu_3^{\it 1}+\nu_4^{\it 1}$  &  $F_2$ &  7346.01&        7346.10& -0.10&  \onlinecite{16ReNiCa.CH4}\\[-1.5mm]
$2\nu_2^{\it 2}+\nu_3^{\it 1}+\nu_4^{\it 1}$  &  $F_2$ &  7365.40&        7365.35&  0.05&  \onlinecite{16ReNiCa.CH4}\\[-1.5mm]
$\nu_1+\nu_2^{\it 1}+\nu_3^{\it 1}$           &  $F_2$ &  7374.25&        7374.42& -0.17&  \onlinecite{16ReNiCa.CH4}\\[-1.5mm]
$4\nu_2^{\it 2}+\nu_4^{\it 1}$                &  $F_2$ &  7384.11&        7384.03&  0.08&  \onlinecite{16ReNiCa.CH4}\\[-1.5mm]
$\nu_2^{\it 1}+2\nu_3^{\it 2}$                &  $F_2$ &  7510.34&        7511.56& -1.22&  \onlinecite{02HiQuxx.CH4}\\[-1.5mm]
$3\nu_2^{\it 1}+\nu_3^{\it 1}$                &  $F_2$ &  7575.86&        7575.43&  0.43&  \onlinecite{16ReNiCa.CH4}\\[-1.5mm]
$3\nu_2^{\it 3}+\nu_3^{\it 1}$                &  $F_2$ &  7584.51&        7583.50&  1.01&  \onlinecite{16ReNiCa.CH4}\\
\hline\hline
\caption*{$^a$ $P_{\mathrm{max}}=14$ value. $^b$ Assigned as $2\nu_1+\nu_4$ in TROVE. $^c$ Value of $7156.72\,$cm$^{-1}$ reported by \citet{14UlBeAl.CH4}.\\}
\end{longtable}

\LTcapwidth=\textwidth
\begin{longtable}{@{\extracolsep{0.2cm}} l c c c r c}
\caption{\label{tab:j0_12ch4_3}Comparison of calculated and experimental $J\!=\!0$ vibrational term values (in cm$^{-1}$) for $^{12}$CH$_4$ in the icosad region and above (see text for a discussion of the experimental uncertainties). The zero-point energy was computed to be $9708.846\,$cm$^{-1}$ at the CVBS limit.}\\ \hline\hline
Mode & Sym. &  Experiment & Calculated & Obs$-$calc & Ref.\\ \hline
\endfirsthead
\caption{(\textit{Continued})}\\ \hline 
Mode & Sym. &  Experiment & Calculated & Obs$-$calc & Ref.\\ \hline
\endhead
$5\nu_4^{\it 5}$                             &  $F_2$ &   6377.53&  6381.09$^{\,a}$& -3.56&  \onlinecite{11NiThRe.CH4}\\[-1.5mm]
$5\nu_4^{\it 1}$                             &  $A_1$ &   6405.89&  6410.06$^{\,a}$& -4.17&  \onlinecite{16ReNiCa.CH4}\\[-1.5mm]
$5\nu_4^{\it 3}$                             &  $F_1$ &   6429.20&          6428.63&  0.57&  \onlinecite{16ReNiCa.CH4}\\[-1.5mm]
$5\nu_4^{\it 3}$                             &  $E$   &   6507.37&          6505.12&  2.25&  \onlinecite{16ReNiCa.CH4}\\[-1.5mm]
$5\nu_4^{\it 5}$                             &  $F_1$ &   6529.74&          6528.34&  1.40&  \onlinecite{16ReNiCa.CH4}\\[-1.5mm]
$\nu_2^{\it 1}+4\nu_4^{\it 0}$               &  $E$   &   6617.50&          6615.81&  1.69&  \onlinecite{16ReNiCa.CH4}\\[-1.5mm]
$\nu_2^{\it 1}+4\nu_4^{\it 2}$               &  $F_1$ &   6638.52&          6636.01&  2.51&  \onlinecite{16ReNiCa.CH4}\\[-1.5mm]
$\nu_2^{\it 1}+4\nu_4^{\it 2}$               &  $A_1$ &   6655.88&          6655.99& -0.11&  \onlinecite{16ReNiCa.CH4}\\[-1.5mm]
$\nu_2^{\it 1}+4\nu_4^{\it 2}$               &  $E$   &   6680.93&          6680.84&  0.09&  \onlinecite{16NiReTa.CH4}\\[-1.5mm]
$\nu_2^{\it 1}+4\nu_4^{\it 4}$               &  $A_2$ &   6682.82&          6681.55&  1.27&  \onlinecite{16ReNiCa.CH4}\\[-1.5mm]
$\nu_2^{\it 1}+4\nu_4^{\it 4}$               &  $F_1$ &   6722.00&          6719.33&  2.67&  \onlinecite{16ReNiCa.CH4}\\[-1.5mm]
$\nu_2^{\it 1}+4\nu_4^{\it 4}$               &  $E$   &   6729.60&          6728.27&  1.33&  \onlinecite{16NiReTa.CH4}\\[-1.5mm]
$\nu_2^{\it 1}+4\nu_4^{\it 4}$               &  $A_1$ &   6737.79&          6737.18&  0.61&  \onlinecite{16ReNiCa.CH4}\\[-1.5mm]
$\nu_2^{\it 1}+4\nu_4^{\it 2}$               &  $A_2$ &   6746.23&          6745.40&  0.83&  \onlinecite{16ReNiCa.CH4}\\[-1.5mm]
$\nu_2^{\it 1}+4\nu_4^{\it 4}$               &  $F_1$ &   6755.38&          6754.15&  1.23&  \onlinecite{16ReNiCa.CH4}\\[-1.5mm]
$\nu_2^{\it 1}+4\nu_4^{\it 4}$               &  $E$   &   6766.23&          6765.13&  1.10&  \onlinecite{16NiReTa.CH4}\\[-1.5mm]
$\nu_1+3\nu_4^{\it 1}$                       &  $A_1$ &   6809.40&          6808.77&  0.63&  \onlinecite{16ReNiCa.CH4}\\[-1.5mm]
$\nu_1+3\nu_4^{\it 3}$                       &  $F_1$ &   6822.30&          6821.92&  0.38&  \onlinecite{16ReNiCa.CH4}\\[-1.5mm]
$\nu_3^{\it 1}+3\nu_4^{\it 1}$               &  $E$   &   6862.74&          6863.53& -0.79&  \onlinecite{16ReNiCa.CH4}\\[-1.5mm]
$\nu_3^{\it 1}+3\nu_4^{\it 1}$               &  $F_1$ &   6862.85&          6863.20& -0.35&  \onlinecite{16NiReTa.CH4}\\[-1.5mm]
$\nu_3^{\it 1}+3\nu_4^{\it 1}$               &  $A_1$ &   6863.10&          6864.32& -1.22&  \onlinecite{16ReNiCa.CH4}\\[-1.5mm]
$2\nu_2^{\it 2}+3\nu_4^{\it 1}$              &  $F_1$ &   6889.68&          6889.53&  0.15&  \onlinecite{16ReNiCa.CH4}\\[-1.5mm]
$2\nu_2^{\it 2}+3\nu_4^{\it 3}$              &  $F_2$ &   6905.60&          6905.65& -0.05&  \onlinecite{16ReNiCa.CH4}\\[-1.5mm]
$\nu_3^{\it 1}+3\nu_4^{\it 3}$               &  $E$   &   6908.80&          6908.84& -0.04&  \onlinecite{16ReNiCa.CH4}\\[-1.5mm]
$\nu_3^{\it 1}+3\nu_4^{\it 3}$               &  $F_1$ &   6915.18&          6915.22& -0.04&  \onlinecite{16ReNiCa.CH4}\\[-1.5mm]
$\nu_3^{\it 1}+3\nu_4^{\it 3}$               &  $A_2$ &   6918.55&          6918.95& -0.40&  \onlinecite{16ReNiCa.CH4}\\[-1.5mm]
$\nu_3^{\it 1}+3\nu_4^{\it 3}$               &  $F_1$ &   6921.58&          6921.75& -0.17&  \onlinecite{16ReNiCa.CH4}\\[-1.5mm]
$\nu_3^{\it 1}+3\nu_4^{\it 3}$               &  $A_1$ &   6922.07&          6923.24& -1.17&  \onlinecite{16ReNiCa.CH4}\\[-1.5mm]
$\nu_3^{\it 1}+3\nu_4^{\it 3}$               &  $E$   &   6925.67&          6927.00& -1.33&  \onlinecite{16ReNiCa.CH4}\\[-1.5mm]
$2\nu_2^{\it 2}+3\nu_4^{\it 1}$              &  $E$   &   6938.40&          6937.71&  0.69&  \onlinecite{16ReNiCa.CH4}\\[-1.5mm]
$2\nu_2^{\it 0}+3\nu_4^{\it 1}$              &  $A_1$ &   6940.10&          6939.47&  0.63&  \onlinecite{16ReNiCa.CH4}\\[-1.5mm]
$2\nu_2^{\it 0}+3\nu_4^{\it 3}$              &  $F_1$ &   6945.16&          6944.87&  0.29&  \onlinecite{16NiReTa.CH4}\\[-1.5mm]
$2\nu_2^{\it 2}+3\nu_4^{\it 3}$              &  $F_1$ &   6949.70&          6949.57&  0.13&  \onlinecite{16ReNiCa.CH4}\\[-1.5mm]
$2\nu_2^{\it 0}+3\nu_4^{\it 3}$              &  $F_2$ &   6962.42&          6962.61& -0.19&  \onlinecite{16ReNiCa.CH4}\\[-1.5mm]
$\nu_1+\nu_2^{\it 1}+2\nu_4^{\it 0}$         &  $E$   &   6990.01&          6990.06& -0.05&  \onlinecite{16ReNiCa.CH4}\\[-1.5mm]
$\nu_1+\nu_2^{\it 1}+2\nu_4^{\it 2}$         &  $F_1$ &   7020.43&          7020.19&  0.24&  \onlinecite{16ReNiCa.CH4}\\[-1.5mm]
$\nu_1+\nu_2^{\it 1}+2\nu_4^{\it 2}$         &  $A_1$ &   7024.03&          7024.05& -0.02&  \onlinecite{16ReNiCa.CH4}\\[-1.5mm]
$\nu_1+\nu_2^{\it 1}+2\nu_4^{\it 2}$         &  $E$   &   7045.69&          7045.89& -0.20&  \onlinecite{16ReNiCa.CH4}\\[-1.5mm]
$\nu_1+\nu_2^{\it 0}+2\nu_4^{\it 2}$         &  $A_2$ &   7056.56&          7056.50&  0.06&  \onlinecite{16ReNiCa.CH4}\\[-1.5mm]
$\nu_2^{\it 1}+\nu_3^{\it 1}+2\nu_4^{\it 0}$ &  $F_1$ &   7085.73&          7085.45&  0.28&  \onlinecite{16ReNiCa.CH4}\\[-1.5mm]
$\nu_2^{\it 1}+\nu_3^{\it 1}+2\nu_4^{\it 0}$ &  $E$   &   7107.28&          7107.39& -0.11&  \onlinecite{16ReNiCa.CH4}\\[-1.5mm]
$\nu_2^{\it 1}+\nu_3^{\it 1}+2\nu_4^{\it 2}$ &  $A_2$ &   7114.54&          7114.43&  0.11&  \onlinecite{16ReNiCa.CH4}\\[-1.5mm]
$3\nu_2^{\it 1}+2\nu_4^{\it 0}$              &  $E$   &   7118.40&          7118.32&  0.08&  \onlinecite{16ReNiCa.CH4}\\[-1.5mm]
$3\nu_2^{\it 3}+2\nu_4^{\it 0}$              &  $A_1$ &   7120.74&          7120.58&  0.16&  \onlinecite{16ReNiCa.CH4}\\[-1.5mm]
$\nu_2^{\it 1}+\nu_3^{\it 1}+2\nu_4^{\it 2}$ &  $F_2$ &   7121.90&          7122.10& -0.20&  \onlinecite{16ReNiCa.CH4}\\[-1.5mm]
$\nu_2^{\it 1}+\nu_3^{\it 1}+2\nu_4^{\it 2}$ &  $F_1$ &   7130.90&          7131.40& -0.50&  \onlinecite{16ReNiCa.CH4}\\[-1.5mm]
$\nu_2^{\it 1}+\nu_3^{\it 1}+2\nu_4^{\it 2}$ &  $A_1$ &   7132.50&          7132.71& -0.21&  \onlinecite{16ReNiCa.CH4}\\[-1.5mm]
$3\nu_2^{\it 3}+2\nu_4^{\it 0}$              &  $A_2$ &   7133.69&          7133.51&  0.18&  \onlinecite{16ReNiCa.CH4}\\[-1.5mm]
$\nu_2^{\it 1}+\nu_3^{\it 1}+2\nu_4^{\it 2}$ &  $E$   &   7134.00&          7134.10& -0.10&  \onlinecite{16ReNiCa.CH4}\\[-1.5mm]
$\nu_2^{\it 1}+\nu_3^{\it 1}+2\nu_4^{\it 2}$ &  $F_1$ &   7139.23&          7140.33& -1.10&  \onlinecite{16ReNiCa.CH4}\\[-1.5mm]
$\nu_2^{\it 1}+\nu_3^{\it 1}+2\nu_4^{\it 2}$ &  $F_2$ &   7141.50&          7142.22& -0.72&  \onlinecite{16ReNiCa.CH4}\\[-1.5mm]
$\nu_2^{\it 1}+\nu_3^{\it 1}+2\nu_4^{\it 2}$ &  $F_1$ &   7151.02&          7151.08& -0.06&  \onlinecite{16ReNiCa.CH4}\\[-1.5mm]
$3\nu_2^{\it 1}+2\nu_4^{\it 2}$              &  $F_1$ &   7153.84&          7153.86& -0.02&  \onlinecite{16ReNiCa.CH4}\\[-1.5mm]
$\nu_1+\nu_3^{\it 1}+\nu_4^{\it 1}$          &  $A_1$ &   7157.16&          7158.06& -0.90&  \onlinecite{16ReNiCa.CH4}\\[-1.5mm]
$\nu_1+\nu_3^{\it 1}+\nu_4^{\it 1}$          &  $E$   &   7164.60&          7165.63& -1.03&  \onlinecite{16ReNiCa.CH4}\\[-1.5mm]
$\nu_1+\nu_3^{\it 1}+\nu_4^{\it 1}$          &  $F_1$ &   7165.60&          7167.95& -2.35&  \onlinecite{16ReNiCa.CH4}\\[-1.5mm]
$3\nu_2^{\it 3}+2\nu_4^{\it 2}$              &  $E$   &   7168.00&          7168.62& -0.62&  \onlinecite{16ReNiCa.CH4}$^{\,b}$\\[-1.5mm]
$3\nu_2^{\it 1}+2\nu_4^{\it 2}$              &  $A_1$ &   7176.10&          7176.09&  0.01&  \onlinecite{16ReNiCa.CH4}\\[-1.5mm]
$3\nu_2^{\it 3}+2\nu_4^{\it 2}$              &  $F_1$ &   7180.00&          7180.01& -0.01&  \onlinecite{16ReNiCa.CH4}\\[-1.5mm]
$3\nu_2^{\it 1}+2\nu_4^{\it 2}$              &  $F_2$ &   7191.05&          7191.12& -0.07&  \onlinecite{16ReNiCa.CH4}\\[-1.5mm]
$3\nu_2^{\it 3}+2\nu_4^{\it 2}$              &  $E$   &   7191.85&          7191.45&  0.40&  \onlinecite{16ReNiCa.CH4}\\[-1.5mm]
$3\nu_2^{\it 1}+2\nu_4^{\it 2}$              &  $E$   &   7217.40&          7217.22&  0.18&  \onlinecite{16ReNiCa.CH4}\\[-1.5mm]
$3\nu_2^{\it 1}+2\nu_4^{\it 2}$              &  $A_2$ &   7221.10&          7220.74&  0.36&  \onlinecite{16ReNiCa.CH4}\\[-1.5mm]
$\nu_1+2\nu_2^{\it 2}+\nu_4^{\it 1}$         &  $F_1$ &   7246.01&          7245.65&  0.36&  \onlinecite{16ReNiCa.CH4}\\[-1.5mm]
$2\nu_1+\nu_2^{\it 1}$                       &  $E$   &   7295.20&          7296.34& -1.14&  \onlinecite{16ReNiCa.CH4}\\[-1.5mm]
$2\nu_3^{\it 2}+\nu_4^{\it 1}$               &  $E$   &   7295.50&          7298.40& -2.90&  \onlinecite{16ReNiCa.CH4}\\[-1.5mm]
$2\nu_3^{\it 2}+\nu_4^{\it 1}$               &  $F_1$ &   7295.80&          7297.66& -1.86&  \onlinecite{16ReNiCa.CH4}\\[-1.5mm]
$2\nu_3^{\it 2}+\nu_4^{\it 1}$               &  $A_1$ &   7299.45&          7300.32& -0.87&  \onlinecite{16ReNiCa.CH4}\\[-1.5mm]
$2\nu_2^{\it 2}+\nu_3^{\it 1}+\nu_4^{\it 1}$ &  $F_1$ &   7326.25&          7326.94& -0.69&  \onlinecite{16ReNiCa.CH4}\\[-1.5mm]
$2\nu_3^{\it 2}+\nu_4^{\it 1}$               &  $F_2$ &   7337.55&          7339.75& -2.20&  \onlinecite{16ReNiCa.CH4}\\[-1.5mm]
$2\nu_3^{\it 2}+\nu_4^{\it 1}$               &  $F_1$ &   7338.16&          7340.03& -1.87&  \onlinecite{16ReNiCa.CH4}\\[-1.5mm]
$2\nu_2^{\it 0}+\nu_3^{\it 1}+\nu_4^{\it 1}$ &  $A_1$ &   7341.60&          7341.87& -0.27&  \onlinecite{16ReNiCa.CH4}\\[-1.5mm]
$2\nu_2^{\it 2}+\nu_3^{\it 1}+\nu_4^{\it 1}$ &  $E$   &   7342.10&          7342.38& -0.28&  \onlinecite{16ReNiCa.CH4}\\[-1.5mm]
$2\nu_2^{\it 2}+\nu_3^{\it 1}+\nu_4^{\it 1}$ &  $F_1$ &   7346.46&          7346.66& -0.20&  \onlinecite{16ReNiCa.CH4}\\[-1.5mm]
$2\nu_2^{\it 2}+\nu_3^{\it 1}+\nu_4^{\it 1}$ &  $A_2$ &   7348.85&          7349.29& -0.44&  \onlinecite{16ReNiCa.CH4}\\[-1.5mm]
$2\nu_2^{\it 2}+\nu_3^{\it 1}+\nu_4^{\it 1}$ &  $E$   &   7352.20&          7352.48& -0.28&  \onlinecite{16ReNiCa.CH4}\\[-1.5mm]
$2\nu_2^{\it 2}+\nu_3^{\it 1}+\nu_4^{\it 1}$ &  $A_1$ &   7360.80&          7361.31& -0.51&  \onlinecite{16ReNiCa.CH4}\\[-1.5mm]
$2\nu_2^{\it 0}+\nu_3^{\it 1}+\nu_4^{\it 1}$ &  $F_1$ &   7368.88&          7368.97& -0.09&  \onlinecite{16ReNiCa.CH4}\\[-1.5mm]
$\nu_1+\nu_2^{\it 1}+\nu_3^{\it 1}$          &  $F_1$ &   7373.16&          7373.97& -0.81&  \onlinecite{16ReNiCa.CH4}\\[-1.5mm]
$4\nu_2^{\it 2}+\nu_4^{\it 1}$               &  $F_1$ &   7394.20&          7393.64&  0.56&  \onlinecite{16ReNiCa.CH4}\\[-1.5mm]
$4\nu_2^{\it 4}+\nu_4^{\it 1}$               &  $F_2$ &   7408.20&          7407.40&  0.80&  \onlinecite{16ReNiCa.CH4}\\[-1.5mm]
$4\nu_2^{\it 4}+\nu_4^{\it 1}$               &  $F_1$ &   7422.30&          7421.35&  0.95&  \onlinecite{16ReNiCa.CH4}\\[-1.5mm]
$4\nu_2^{\it 2}+\nu_4^{\it 1}$               &  $F_2$ &   7436.30&          7435.90&  0.40&  \onlinecite{16ReNiCa.CH4}\\[-1.5mm]
$\nu_1+3\nu_2^{\it 1}$                       &  $E$   &   7447.52&          7447.83& -0.31&  \onlinecite{16ReNiCa.CH4}\\[-1.5mm]
$\nu_1+3\nu_2^{\it 3}$                       &  $A_2$ &   7468.21&          7467.33&  0.88&  \onlinecite{16ReNiCa.CH4}\\[-1.5mm]
$\nu_1+3\nu_2^{\it 3}$                       &  $A_1$ &   7468.50&          7467.42&  1.08&  \onlinecite{16ReNiCa.CH4}\\[-1.5mm]
$\nu_2^{\it 1}+2\nu_3^{\it 0}$               &  $E$   &   7483.67&          7483.79& -0.12&  \onlinecite{16ReNiCa.CH4}\\[-1.5mm]
$\nu_2^{\it 1}+2\nu_3^{\it 2}$               &  $F_1$ &   7512.26&          7513.39& -1.13&  \onlinecite{16ReNiCa.CH4}\\[-1.5mm]
$\nu_2^{\it 1}+2\nu_3^{\it 2}$               &  $E$   &   7552.23&          7553.79& -1.56&  \onlinecite{16ReNiCa.CH4}\\[-1.5mm]
$\nu_2^{\it 1}+2\nu_3^{\it 2}$               &  $A_1$ &   7559.00&          7560.60& -1.60&  \onlinecite{16ReNiCa.CH4}\\[-1.5mm]
$3\nu_2^{\it 1}+\nu_3^{\it 1}$               &  $F_1$ &   7569.51&          7569.25&  0.26&  \onlinecite{16ReNiCa.CH4}\\[-1.5mm]
$3\nu_2^{\it 3}+\nu_3^{\it 1}$               &  $F_1$ &   7580.90&          7580.36&  0.54&  \onlinecite{16ReNiCa.CH4}\\[-1.5mm]
$2\nu_1+2\nu_4^{\it 2}$                      &  $F_2$ &   8388.00&          8384.52&  3.48&  \onlinecite{14UlBeAl.CH4}\\[-1.5mm]
$\nu_1+\nu_3^{\it 1}+2\nu_4^{\it 2}$         &  $F_2$ &   8421.00&          8422.37& -1.37&  \onlinecite{14UlBeAl.CH4}\\[-1.5mm]
$\nu_1+2\nu_3^{\it 2}$                       &  $F_2$ &   8618.67&          8613.92&  4.75&  \onlinecite{14UlBeAl.CH4}\\[-1.5mm]
$2\nu_1+\nu_3$                               &  $F_2$ &   8808.95&   8812.01$^{\,a,c}$& -3.06&  \onlinecite{14UlBeAl.CH4}\\[-1.5mm]
$3\nu_3^{\it 1}$                             &  $F_2$ &   8907.30&          8909.59& -2.29&  \onlinecite{14UlBeAl.CH4}\\[-1.5mm]
$3\nu_3^{\it 3}$                             &  $F_2$ &   9045.96&          9048.87& -2.91&  \onlinecite{14UlBeAl.CH4}\\[-1.5mm]
$\nu_1+2\nu_3^{\it 0}+\nu_4^{\it 1}$         &  $F_2$ &   9888.47&   9892.46$^{\,a}$& -3.99&  \onlinecite{14UlBeAl.CH4}\\[-1.5mm]
$\nu_1+\nu_2+2\nu_3$                         &  $F_2$ &  10115.67&   $^{\,d}$      &   {--} &  \onlinecite{14UlBeAl.CH4}\\[-1.5mm]
$3\nu_3+\nu_4$                               &  $F_2$ &  10265.59&   $^{\,d}$      &   {--} &  \onlinecite{14UlBeAl.CH4}\\[-1.5mm]
$2\nu_1+\nu_2+\nu_3$                         &  $F_2$ &  10302.17&   $^{\,d}$      &   {--} &  \onlinecite{14UlBeAl.CH4}\\[-1.5mm]
$\nu_1+3\nu_3$                               &  $F_2$ &  11276.31&         11277.96$^{\,c}$& -1.65&  \onlinecite{14UlBeAl.CH4}\\
\hline\hline
\caption*{$^a$ $P_{\mathrm{max}}=14$ value. $^b$ Assigned as $\nu_1+\nu_3+\nu_4$ in TROVE. $^{\,c}$ Unable to identify vibrational angular momentum quantum numbers. $^{\,d}$ Unable to identify energy level in TROVE.\\}
\end{longtable}

\subsection{Equilibrium geometry and pure rotational energies}
\label{sec:eq_rotational}

 The value of $r_{\mathrm{ref}}$ used in Eq.~\eqref{eq:stretch} does not define the minimum of the PES because a linear expansion term has been included in the parameter set. The true equilibrium C{--}H bond length determined from the CBS-F12$^{\,\mathrm{HL}}$ PES is listed in Table~\ref{tab:eq_ref}. It is in excellent agreement with previous values which is gratifying as it has been calculated in a purely \textit{ab initio} fashion.

\begin{table}
\tabcolsep=0.35cm
\caption{\label{tab:eq_ref}Equilibrium C{--}H bond length}
\begin{center}
\begin{tabular}{l c c}
\hline\hline
\multicolumn{1}{c}{$r$(C{--}H) / $\mathrm{\AA}$} & Ref. & \multicolumn{1}{c}{Approach}\\
\hline
1.08601     & This work & Purely \textit{ab initio} PES\\[-1.5mm]
1.08598     & This work & Refined geometry PES\\[-1.5mm]
1.08601(4)  & \onlinecite{11NiReTy.CH4} & Empirically adjusted PES\\[-1.5mm]
1.08609     & \onlinecite{14WaCaxx.CH4} & Empirically refined PES\\[-1.5mm]
1.08595(30) & \onlinecite{99Stxxxx.CH4} & Combined experimental and \textit{ab initio} analysis\\[-1.5mm]
1.086(2) & \onlinecite{94HoMaQu.CH4} & Quantum Monte Carlo calculations\\[-1.5mm]
1.0847 & \onlinecite{09AlBaBo.CH4} & Effective Hamiltonian model\\[-1.5mm]
1.08553(4) & \onlinecite{16AmLoPi.CH4} & Effective Hamiltonian model\\
\hline\hline
\end{tabular}
\end{center}
\end{table}

 However, it is more informative to look at pure rotational energies as these are highly dependent on the molecular geometry through the moments of inertia. In Table~\ref{tab:rotational}, computed rotational energy levels up to $J\!=\!10$ are compared against experimental values listed in \citet{11NiReTy.CH4} (originally attributed to the spherical top data system,~\citep{STDS:1998} which contains measurements from \citet{85OlAnBa.CH4}). Calculations were carried out with $P_{\mathrm{max}}=12$ which is sufficient for converging ground state rotational energies.

\LTcapwidth=\textwidth
\begin{longtable}{@{\extracolsep{0.1cm}} c c c c c c r r}
\caption{\label{tab:rotational}Comparison of calculated and experimental $J\leq10$ pure rotational energy levels (in cm$^{-1}$) for $^{12}$CH$_4$. The experimental ground state values are from \citet{11NiReTy.CH4} but are originally attributed to the spherical top data system.~\citep{STDS:1998} Computed values correspond to the \textit{ab initio} geometry (A) and the empirically refined geometry (B) (see text).}\\ \hline\hline
$J$ & $K$ & Sym. &  Experiment & Calculated (A) & Calculated (B) & Obs$-$calc (A) & Obs$-$calc (B) \\ \hline
\endfirsthead
\caption{(\textit{Continued})}\\ \hline 
$J$ & $K$ & Sym. &  Experiment & Calculated (A) & Calculated (B) & Obs$-$calc (A) & Obs$-$calc (B) \\ \hline
\endhead
  0&   0& $A_1$ &    0.00000&    0.00000&    0.00000&  0.00000&  0.00000\\[-1.5mm]
  1&   1& $F_1$ &   10.48165&   10.48105&   10.48164&  0.00060&  0.00001\\[-1.5mm]
  2&   1& $F_2$ &   31.44239&   31.44061&   31.44235&  0.00178&  0.00004\\[-1.5mm]
  2&   2& $E$   &   31.44212&   31.44034&   31.44209&  0.00178&  0.00003\\[-1.5mm]
  3&   1& $F_2$ &   62.87684&   62.87329&   62.87678&  0.00355&  0.00006\\[-1.5mm]
  3&   2& $A_2$ &   62.87817&   62.87462&   62.87811&  0.00355&  0.00006\\[-1.5mm]
  3&   3& $F_1$ &   62.87578&   62.87222&   62.87571&  0.00356&  0.00007\\[-1.5mm]
  4&   0& $A_1$ &  104.77284&  104.76692&  104.77274&  0.00592&  0.00010\\[-1.5mm]
  4&   1& $F_1$ &  104.77470&  104.76879&  104.77460&  0.00591&  0.00010\\[-1.5mm]
  4&   2& $E$   &  104.77603&  104.77012&  104.77594&  0.00591&  0.00009\\[-1.5mm]
  4&   3& $F_2$ &  104.78001&  104.77411&  104.77993&  0.00590&  0.00008\\[-1.5mm]
  5&   1& $F_1$ &  157.12434&  157.11548&  157.12420&  0.00886&  0.00014\\[-1.5mm]
  5&   2& $E$   &  157.13719&  157.12837&  157.13709&  0.00882&  0.00010\\[-1.5mm]
  5&   3& $F_1$ &  157.13892&  157.13010&  157.13882&  0.00882&  0.00010\\[-1.5mm]
  5&   5& $F_2$ &  157.12793&  157.11908&  157.12780&  0.00885&  0.00013\\[-1.5mm]
  6&   1& $F_2$ &  219.91505&  219.90268&  219.91487&  0.01237&  0.00018\\[-1.5mm]
  6&   2& $A_2$ &  219.91985&  219.90750&  219.91969&  0.01235&  0.00016\\[-1.5mm]
  6&   3& $F_1$ &  219.94126&  219.92897&  219.94117&  0.01229&  0.00009\\[-1.5mm]
  6&   4& $A_1$ &  219.94523&  219.93295&  219.94515&  0.01228&  0.00008\\[-1.5mm]
  6&   5& $F_2$ &  219.93677&  219.92446&  219.93666&  0.01231&  0.00011\\[-1.5mm]
  6&   6& $E$   &  219.91346&  219.90109&  219.91328&  0.01237&  0.00018\\[-1.5mm]
  7&   1& $F_1$ &  293.12299&  293.10652&  293.12277&  0.01647&  0.00022\\[-1.5mm]
  7&   1& $F_2$ &  293.12655&  293.11010&  293.12634&  0.01645&  0.00021\\[-1.5mm]
  7&   2& $A_2$ &  293.15420&  293.13783&  293.15408&  0.01637&  0.00012\\[-1.5mm]
  7&   3& $F_2$ &  293.16457&  293.14823&  293.16448&  0.01634&  0.00009\\[-1.5mm]
  7&   5& $F_1$ &  293.17868&  293.16238&  293.17864&  0.01630&  0.00004\\[-1.5mm]
  7&   6& $E$   &  293.17013&  293.15381&  293.17007&  0.01632&  0.00006\\[-1.5mm]
  8&   0& $A_1$ &  376.73044&  376.70932&  376.73019&  0.02112&  0.00025\\[-1.5mm]
  8&   1& $F_1$ &  376.73372&  376.71261&  376.73349&  0.02111&  0.00023\\[-1.5mm]
  8&   2& $E$   &  376.82129&  376.80044&  376.82133&  0.02085& -0.00004\\[-1.5mm]
  8&   3& $F_1$ &  376.80478&  376.78388&  376.80476&  0.02090&  0.00002\\[-1.5mm]
  8&   3& $F_2$ &  376.82627&  376.80544&  376.82632&  0.02083& -0.00005\\[-1.5mm]
  8&   5& $F_2$ &  376.78587&  376.76492&  376.78581&  0.02095&  0.00006\\[-1.5mm]
  8&   6& $E$   &  376.73565&  376.71454&  376.73541&  0.02111&  0.00024\\[-1.5mm]
  9&   1& $F_1$ &  470.71696&  470.69064&  470.71670&  0.02632&  0.00026\\[-1.5mm]
  9&   1& $F_2$ &  470.72034&  470.69403&  470.72009&  0.02631&  0.00025\\[-1.5mm]
  9&   2& $E$   &  470.79897&  470.77290&  470.79898&  0.02607& -0.00001\\[-1.5mm]
  9&   3& $F_1$ &  470.80528&  470.77923&  470.80531&  0.02605& -0.00003\\[-1.5mm]
  9&   4& $A_1$ &  470.83096&  470.80498&  470.83106&  0.02598& -0.00010\\[-1.5mm]
  9&   5& $F_2$ &  470.86506&  470.83918&  470.86528&  0.02588& -0.00022\\[-1.5mm]
  9&   6& $A_2$ &  470.87292&  470.84707&  470.87315&  0.02585& -0.00023\\[-1.5mm]
  9&   7& $F_1$ &  470.85500&  470.82910&  470.85517&  0.02590& -0.00017\\[-1.5mm]
 10&   1& $F_1$ &  575.18430&  575.15264&  575.18447&  0.03166& -0.00017\\[-1.5mm]
 10&   1& $F_2$ &  575.05266&  575.02059&  575.05242&  0.03207&  0.00024\\[-1.5mm]
 10&   2& $A_2$ &  575.05567&  575.02361&  575.05544&  0.03206&  0.00023\\[-1.5mm]
 10&   3& $F_2$ &  575.17008&  575.13837&  575.17019&  0.03171& -0.00011\\[-1.5mm]
 10&   5& $F_1$ &  575.25978&  575.22834&  575.26020&  0.03144& -0.00042\\[-1.5mm]
 10&   6& $E$   &  575.27192&  575.24050&  575.27236&  0.03142& -0.00044\\[-1.5mm]
 10&   7& $F_2$ &  575.28542&  575.25405&  575.28589&  0.03137& -0.00047\\[-1.5mm]
 10&   8& $A_1$ &  575.22292&  575.19137&  575.22321&  0.03155& -0.00029\\[-1.5mm]
 10&  10& $E$   &  575.05127&  575.01920&  575.05101&  0.03207&  0.00026\\
\hline\hline
\end{longtable}
 
 The CBS-F12$^{\,\mathrm{HL}}$ PES consistently underestimates ground state rotational energy levels and the residual error increases systematically by about $0.00060\,$cm$^{-1}$ at each step up in $J$. Overall, the $51$ energies are reproduced with a rms error of $0.02008\,$cm$^{-1}$. This is around two orders of magnitude larger than the empirically adjusted PES of \citet{11NiReTy.CH4} which yields an identical value of  $r$(C{--}H)$=1.08601\,\mathrm{\AA}$ for the C{--}H bond length but a rms error of $0.00029\,$cm$^{-1}$.
 
\begin{figure}
\includegraphics{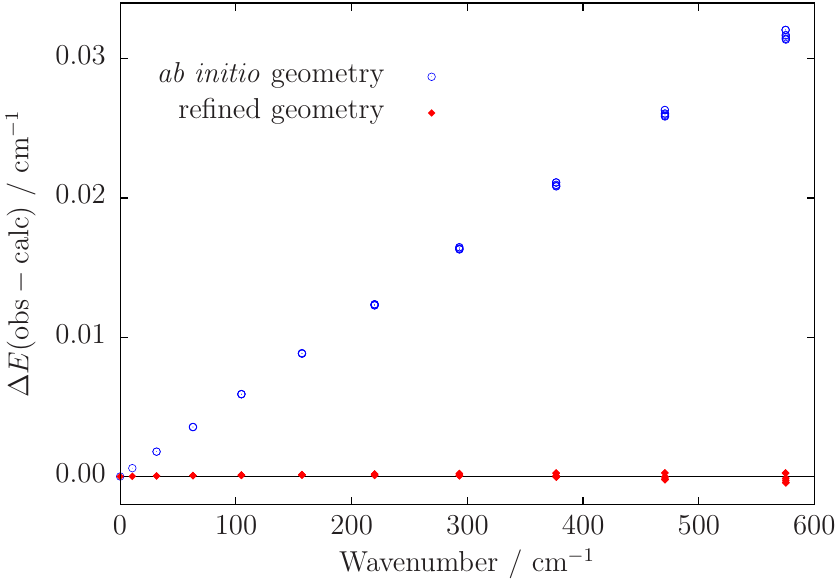}
\caption{\label{fig:res_rot}Residual errors $\Delta E(\mathrm{obs}-\mathrm{calc})$ for computed pure rotational energies using the \textit{ab initio} and empirically refined equilibrium geometry (see Table~\ref{tab:rotational}).}
\end{figure}

 To help explain this discrepancy it is relatively straightforward to improve the CBS-F12$^{\,\mathrm{HL}}$ results by refining the equilibrium geometry. This is done through a nonlinear least-squares fitting to the experimental energy levels and can significantly improve the accuracy of computed intra-band rotational wavenumbers.~\cite{YuBaYa09.NH3,13YaPoTh.H2CS,15OwYuYa.SiH4} After two iterations refining the parameter $r_{\mathrm{ref}}$, the experimental energy levels up to $J=10$ are reproduced with a rms error of $0.00018\,$cm$^{-1}$ (see Table~\ref{tab:rotational} and Fig.~\ref{fig:res_rot}) and this corresponds to a bond length of $r$(C{--}H)$=1.08598\,\mathrm{\AA}$ (also given in Table~\ref{tab:eq_ref}). This value is within the uncertainty of the bond length from \citet{11NiReTy.CH4} and is remarkably close to the original \textit{ab initio} result. However, we have refrained from adopting the new equilibrium geometry for the CBS-F12$^{\,\mathrm{HL}}$ PES as it leads to a poorer description of vibrational energies (see for example Ref.~\onlinecite{15OwYuYa.SiH4}), which were the main focus of this work.
 
\section{Conclusions}
\label{sec:conc}

 State-of-the-art electronic structure calculations have been used to generate a new nine-dimensional PES for methane. The CBS-F12$^{\,\mathrm{HL}}$ PES represents the most accurate \textit{ab initio} surface to date. This is confirmed by the achievement of sub-wavenumber accuracy for a considerable number of vibrational energy levels including those at higher energies. Although the computed \textit{ab initio} equilibrium C{--}H bond length was in excellent agreement with previous values, systematic errors arose in calculated pure rotational energies of $^{12}$CH$_4$. These errors could be significantly reduced by adjusting the equilibrium geometry of the CBS-F12$^{\,\mathrm{HL}}$ PES. The resultant bond length was remarkably close to the original \textit{ab initio} value and still consistent with prior studies.
 
  Despite the advances in electronic structure theory the best \textit{ab initio} PES is rarely accurate enough for the requirements of high-resolution spectroscopy and empirical refinement is a necessary step. Refinement can be a computationally intensive process~\cite{YuBaTe11.NH3} but it can produce orders-of-magnitude improvements in the accuracy of computed rovibrational energy levels. It is natural then to question the benefit of using sophisticated methods with large basis sets to generate the original \textit{ab initio} surface. Whilst a better \textit{ab initio} PES will lead to a superior refinement, at some stage the gain in accuracy when simulating rotation-vibration spectra will not correlate with the computational cost of improving the underlying \textit{ab initio} surface. For this reason we believe that more sophisticated electronic structure calculations to improve the CBS-F12$^{\,\mathrm{HL}}$ PES are currently not worthwhile. The CBS-F12$^{\,\mathrm{HL}}$ PES will serve as an excellent starting point for refinement and we recommend this surface for future use.
  
\section*{Supplementary Material}

See supplementary material for the expansion parameters and corresponding program to construct the CBS-F12$^{\,\mathrm{HL}}$ PES. A list of computed vibrational $J\!=\!0$ energy levels of $^{12}$CH$_4$ is also provided.

\begin{acknowledgments}
This work was supported by ERC Advanced Investigator Project 267219, and FP7-MC-IEF project 629237.
\end{acknowledgments}

\bibliography{ch4_AO}

\end{document}